\documentclass[twocolumn,superscriptaddress]{revtex4-1}

\usepackage{amsfonts,amsmath,amssymb}
\usepackage{graphicx}
\usepackage[usenames,dvipsnames]{xcolor}

\usepackage{lettrine}
\renewcommand{\figurename}{\textbf{Figure}}
\def\fnum@figure{\figurename\nobreakspace\textbf{\thefigure}}

\def \b{\hat{\beta}}
\def \bd{\hat{\beta}^\dag}
\def \ao{\hat{a}_1}
\def \aod{\hat{a}_1^\dag}
\def \at{\hat{a}_2}
\def \atd{\hat{a}_2^\dag}
\def \Eb{E_\beta}

\def \nthM{\bar{n}^\mathrm{th}_{M}}
\def \nthb{\bar{n}^\mathrm{th}_{\beta}}
\def \gts{g^{(2)}_{a_s}(0)}

\newcommand{\as}[1]{\hat{a}_\mathrm{s}}
\newcommand{\As}[1]{\hat{a}_\mathrm{s}^\dag}
\newcommand{\xzpf}{x_\mathrm{zpf}}
\newcommand{\DeltaM}{\Delta}
\newcommand{\Deltaot}{\delta}
\newcommand{\blank}{\vspace{3mm}\noindent}

\begin{document}

\title{Enhanced nonlinear interactions in quantum optomechanics via mechanical amplification}
\author{Marc-Antoine Lemonde}
\affiliation{Department of Physics, McGill University, 3600 rue University, Montreal, Quebec H3A 2T8, Canada}
\author{Nicolas Didier}
\affiliation{Department of Physics, McGill University, 3600 rue University, Montreal, Quebec H3A 2T8, Canada}
\affiliation{D\'epartment de Physique, Universit\'e de Sherbrooke, 2500 boulevard de l'Universit\'e, Sherbrooke, Qu\'ebec J1K 2R1, Canada}
\author{Aashish~A.~Clerk}
\affiliation{Department of Physics, McGill University, 3600 rue University, Montreal, Quebec H3A 2T8, Canada}

\begin{abstract}
We present an approach for exponentially enhancing the single-photon coupling strength in an optomechanical system using only additional linear resources.  It allows one to
reach the quantum nonlinear regime of optomechanics, where nonlinear effects are observed at the single photon level, even if the bare coupling strength is much smaller than the mechanical frequency and cavity damping rate.  
Our method is based on using a large amplitude, strongly detuned mechanical parametric drive to amplify mechanical zero-point fluctuations and hence enhance the radiation pressure interaction.
It has the further benefit of allowing time-dependent control, enabling pulsed schemes.  
For a two-cavity optomechanical setup, we show that our scheme generates photon blockade for experimentally accessible parameters, and even makes the production
of photonic states with negative Wigner functions possible.
We discuss how our method is an example of a more general strategy for enhancing boson-mediated two-particle interactions and nonlinearities.
\clearpage\end{abstract}

\maketitle

%


\lettrine{T}{he} field of quantum cavity optomechanics aims at synthesizing quantum states of light and motion using radiation pressure, the fundamental nonlinear interaction between photons and phonons.
Considerable effort is currently devoted to reaching the true quantum regime, where nonlinear signatures are observed at the single-photon level~\cite{OptomechanicsBook, OptomechanicsRMP}. 
In the canonical system of a cavity comprising a movable mirror, the quantum nonlinear regime requires the single-photon
coupling constant~$g$ to be  
comparable to both the mechanical resonator frequency $\omega_M$, 
as well as the cavity damping rate $\kappa$~\cite{Rabl_PRL_2011, Nunnenkamp_PRL_2011,Kronwald_PRA_2013,Kronwald_PRL_2013}.  Current experiments are still far from this regime. 

The simplest strategy to enhance the optomechanical interaction is to coherently drive the cavity.
This approach has facilitated a wide variety of interesting phenomena, ranging from ground-state cooling of the mechanical resonator~\cite{Teufel_Nature_2011,Chan_Nature_2011} 
to mechanically-mediated state transfer~\cite{Palomaki_2013} and the generation of squeezed light~\cite{Brooks_Nature_2012,Safavi-Naeini_Nature_2013,Regal2013}.  The optomechanical interaction is however effectively linearized in this strong driving regime, and hence
there is generally no enhancement of quantum nonlinear effects.
For enhanced nonlinearity, one can tune the strong drive so that the weak residual optomechanical nonlinearity becomes resonant~\cite{Lemonde_PRL_2013, Lemonde_PRA_2015}.
The quantum regime is then reached for $g \sim \kappa$, where the damping rate of the cavity $\kappa$ can be much smaller than $\omega_M$. 
A similar enhancement of quantum nonlinear effects is found in undriven two-cavity setups, where the energy difference between the optical modes is set to render the nonlinear optomechanical interaction resonant~\cite{Komar_Bennett_PRA_2013} or nearly-resonant~\cite{Ludwig_Safavi_PRL_2012,Liao_Nori_2015}.
Enhancement of the nonlinearity has also been proposed in a transient scheme~\cite{Xu_PRA_2014}.
Experimentally, these approaches are still not sufficient:  for systems in the optimal good cavity regime ($\omega_M > \kappa$), the largest achieved couplings $g$ are at most a percent of $\kappa$~\cite{OptomechanicsBook, OptomechanicsRMP, Chan_2012}.

In this paper, we present a new method for enhancing the single-photon optomechanical interaction for systems deep in the well-resolved sideband regime.  It enables true quantum nonlinearity
even when the single photon coupling $g$ is much smaller than the cavity damping rate $\kappa$.  
Crucially, our scheme results in a tunable nonlinearity, and only requires additional linear resources:  
it does not require a coupling to an auxiliary quantum nonlinear system (like a qubit~\cite{Armour_2002, Pirkkalainen_2015, Didier_2011}).
The key idea is to use detuned parametric driving of the mechanics
to increase the effective scale of mechanical zero-point position fluctuations $\xzpf$.  
This amplification
directly enhances the coupling strength (as $g\propto\xzpf$), while the large detuning allows the mechanics to still effectively mediate a photon-photon interaction.
So far, parametric mechanical driving has been studied only in the linearized regime of optomechanics~\cite{Szorkovszky_2011,Szorkovszky_2014,Farace_PRA_2012}.

Combined with the resonant enhancement possible in two-cavity setups~\cite{Komar_Bennett_PRA_2013,Ludwig_Safavi_PRL_2012,Liao_Nori_2015}, our novel approach lets one reach the quantum regime in current state-of-the-art experiments ($g \sim 10^{-2} \kappa$).
In addition, by controlling the parametric drive amplitude, the nonlinear interaction can be rapidly turned on and off in time, greatly extending its utility.
We stress that due to the fundamental asymmetry between photons and phonons in the optomechanical interaction, parametrically driving the cavity~\cite{Liu_Nori_PRL_2015} {\it does not} enhance single-photon quantum effects.  
While such photonic parametric driving generates an enhanced nonlinearity, 
this nonlinearity 
necessarily involves states with large photon numbers (i.e. squeezed Fock states), reducing its utility~\cite{SMnatcom}.
As we discuss in detail, parametrically driving the mechanics results in very different physics and a true enhancement of {\it single-photon} nonlinearity.

The approach outlined in our work is a particular example of a general strategy for enhancing two-particle interactions using only linear resources.  It could thus have applications to continuous variable quantum information processing, where strong nonlinearities are crucial for universal control, but often difficult to achieve~\cite{Braunstein_2005}.
In our optomechanical system, the mechanical resonator mediates an effective retarded interaction between photons~\cite{Bose_PRA_1997, Rabl_PRL_2011, Nunnenkamp_PRL_2011, Ludwig_Safavi_PRL_2012}.  Our scheme enhances this interaction by using a parametric drive to
manipulate the mechanical dynamics.
Similar improvements can be obtained in any system where bosonic modes mediate a two-particle interaction:
by parametrically driving the intermediate modes, interactions can be greatly enhanced (see, e.g., phonon-mediated electron-electron interactions in superconductivity~\cite{Hakioglu_1993,Misochko_2011,Misochko_2013}).  
An intuitive picture of the physics is provided by the effective Keldysh action describing the cavity photons in our system.
This approach explicitly connects the nonlinear interaction to the mechanical Green's functions, and shows how a large detuning of the parametric drive is important to get a time-local interaction.

\begin{figure}[t]
\includegraphics[width=0.8\columnwidth]{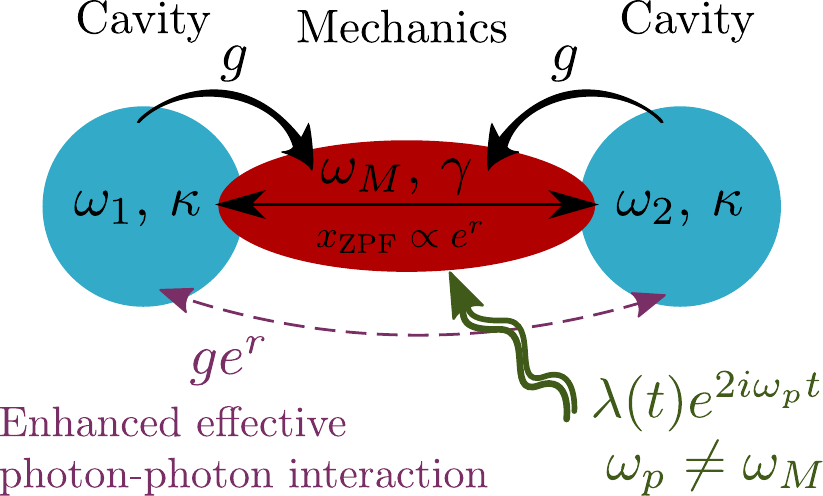}
\caption{\textbf{Sketch of the system.} Tunneling between two optical cavities (blue circles) is mediated by a mechanical mode (red ellipse) on which a large amplitude, strongly detuned parametric drive is applied to amplify its $\hat{X}$ quadrature. This scheme results in an exponential enhancement of the single-photon coupling constant $g$, thereby amplifying the resulting effective photon-photon interaction.
}
\label{Fig:schema}
\end{figure}

\blank
\textbf{Results}\\
\textbf{System.}
We consider an optomechanical (OM) system consisting of two optical modes coupled to a single mechanical resonator (MR) via radiation pressure (cf.~Fig.~\ref{Fig:schema}), where the interaction is of the form $g(\atd\ao + \aod\at)(\hat{b} + \hat{b}^\dag)$.
Here, $\hat{a}_{1,2}$ and $\hat{b}$ are the annihilation operators of the optical modes $1,2$ and the MR, respectively.
Such three-mode OM systems have been discussed extensively in the literature~\cite{Komar_Bennett_PRA_2013, Ludwig_Safavi_PRL_2012, Liao_Nori_2015} and have been realized experimentally~\cite{Thompson_Harris_Nature2008, Grudinin_Vahala_PRL2010, Safavi-Naeini_NJP_2011}.
As already discussed, if one tunes the mode splitting 
$\omega_{21} \equiv \omega_2 - \omega_1$ to make the optomechanical interaction resonant, 
quantum nonlinear effects can be observed when the OM coupling $g$ is comparable to the damping rate $\kappa$ of the cavities~\cite{Komar_Bennett_PRA_2013, Ludwig_Safavi_PRL_2012, Liao_Nori_2015}. 

We wish to enhance this generic system so that single-photon quantum effects are possible even when $g \ll \kappa$.  To that end, we parametrically modulate the MR spring constant at frequency $2\omega_p$ (cf.~Fig.~\ref{Fig:schema})~\cite{Rugar_1991}. The system Hamiltonian then reads
\begin{equation}
\hat{H}= \DeltaM \hat{b}^\dag\hat{b} - \tfrac{1}{2}(\lambda \hat{b}^2 + \lambda^* \hat{b}^{\dag2}) + g[ \atd\ao\hat{b}e^{-i\delta t}+\mathrm{H.c.}].
\label{Eq:Hinitial}
\end{equation}
Here $\delta = \omega_p-\omega_{21}$ is the detuning of the parametric drive frequency from the optical mode splitting.
We work in an interaction picture with respect to the free cavity Hamiltonians and, for the mechanics, with respect to the pump frequency $\omega_p$.
The parameter $\lambda$ (taken to be real) is the parametric drive strength, and $\DeltaM \equiv \omega_M - \omega_p$.
We have assumed $\omega_p + \omega_{21}$ large enough to neglect highly non-resonant interaction terms;
this approximation is always valid for the parameters considered in this work (see supplemental information \cite{SMnatcom}).
In what follows, we always stay in the regime where the MR is stable even without dissipation, i.e.~$\lambda < \DeltaM$.
The quadratic part of $\hat{H}$ is then diagonal when expressed in terms of the Bogoliubov mode $\b$, defined as $\b = \hat{b} \cosh r - \hat{b}^\dag \sinh r$, with energy $\Eb = \DeltaM/\cosh 2r$. The parameter $r$ is set by the parametric drive strength, $\tanh2r = \lambda/\DeltaM$. 
Experimentally, detuned parametric drives have already been employed in optomechanics setups~\cite{Szorkovszky_2013}, and are particularly compatible with recent
state-of-the-art electromechanical setups~\cite{Andrews_2015}.

\blank
\textbf{Enhanced, tunable nonlinear interactions.}
The detuning $\delta$ can be chosen to
select the nature of the nonlinear
interaction that is effectively amplified.
Taking $\Deltaot = 0$ gives rise to the interaction
\begin{equation}
	\hat{H}_\mathrm{SRP}=\Eb\bd\b+\tilde{g}(\atd\ao+\aod\at)( \b+\bd)+\hat{H}_\mathrm{SRP}'.
	\label{H2PT}
\end{equation}
For large amplification (i.e.~for $e^{2r}\gg1$) and a state where the Bogoliubov mode is not strongly squeezed,
the term $\hat{H}_\mathrm{SRP}'=\frac{1}{2}ge^{-r}(\b - \bd)(\atd\ao-\aod\at)$
can be dropped and $\hat{H}_\mathrm{SRP}$ becomes similar to the standard radiation pressure interaction in the two-cavity OM system.  
We now however have an {\it exponentially-enhanced} effective single-photon coupling constant,
\begin{equation}
\tilde{g}=\tfrac{1}{2}ge^r\gg g.
\label{gtilde}
\end{equation}
This enhancement is a direct consequence of the parametric drive: it amplifies the vacuum fluctuations of the mechanical $\hat{X} \equiv \hat{b} + \hat{b}^\dagger$ quadrature, and thus enhances the coupling of the cavities to this quadrature.  The effective photon-photon interaction induced by Eq.~\eqref{H2PT} is further enhanced compared to a standard single-cavity OM setup, as the Bogoliubov mode energy $E_{\beta}$ is also tunable and can be made much smaller than $\omega_M$.
However, one also needs this mechanically-mediated interaction to be sufficiently time-local; as shown below, this further constrains $\Eb > \tilde{g}, \kappa$.
The induced photon-photon interaction thus scales as $\tilde{g}^2 / E_{\beta}$, as opposed to $\sim g^2 / \omega_M$ in a standard OM cavity~\cite{Rabl_PRL_2011, Nunnenkamp_PRL_2011}.
We stress that only the amplification effect of the parametric drive is crucial here.
This means that the mechanics does not have to be in a vacuum squeezed state (i.e.~the Bogoliubov mode can have a thermal population).

If one instead tunes frequencies so that
$\Deltaot \approx \Eb > \tilde{g}, \kappa$, one can make an additional rotating wave approximation, yielding the interaction ($e^{2r} \gg1$)
\begin{equation}
\hat{H}_\mathrm{PAT}=(\Eb-\Deltaot)\bd\b+\tilde{g}(\atd\ao\b+\aod\at\bd).
\label{HPAT}
\end{equation}
This is a phonon-assisted  photon-tunneling interaction, with an enhanced interaction strength $\tilde{g}$ again given by Eq.~(\ref{gtilde}).
This form of interaction (without any parametric enhancement) has been studied in the resonant regime ($\Eb = \Deltaot$)~\cite{Komar_Bennett_PRA_2013}, as well as in the detuned regime ($\tilde{g}, \kappa < \vert  \Eb - \Deltaot \vert \ll \vert  \Eb + \Deltaot \vert$)~\cite{Ludwig_Safavi_PRL_2012}. 
While tuning the parametric drive frequency lets us pick the form of the effective nonlinear interaction, tuning its amplitude lets us control the interaction strength.
As discussed below, the possibility to modulate the interaction strength in time is extremely useful to prepare the $\beta$ mode in the desired state while preventing mechanical heating.

We stress that our scheme allows in principle arbitrarily large nonlinearity enhancements in optomechanics using only additional linear resources (i.e.~a large parametric drive that is strongly detuned). In practice, the achievable enhancement will be limited by the maximum detuning $\Delta$ possible (needed to ensure $E_\beta > \tilde{g}, \kappa$) and by the stability of the parametric drive (i.e.~one should not cross the instability threshold).
In addition, the standard realization of mechanical parametric driving (i.e.~spring constant modulation) results in $\lambda < \omega_M$; 
as $\lambda \sim \Delta$ for large amplifications $r$, this implies $\Delta < \omega_M$.
Thus, if one wants to use large $r$ values to enhance the interaction, the requirement that $\Eb > \kappa$ implies that the system must be deep in the well-resolved sideband regime.
Despite these caveats, our approach represents a practically-attractive route towards single-photon strong coupling, given the difficulty of engineering systems with an intrinsically large value of $g$.

\begin{figure}[t]
\includegraphics[width=\columnwidth]{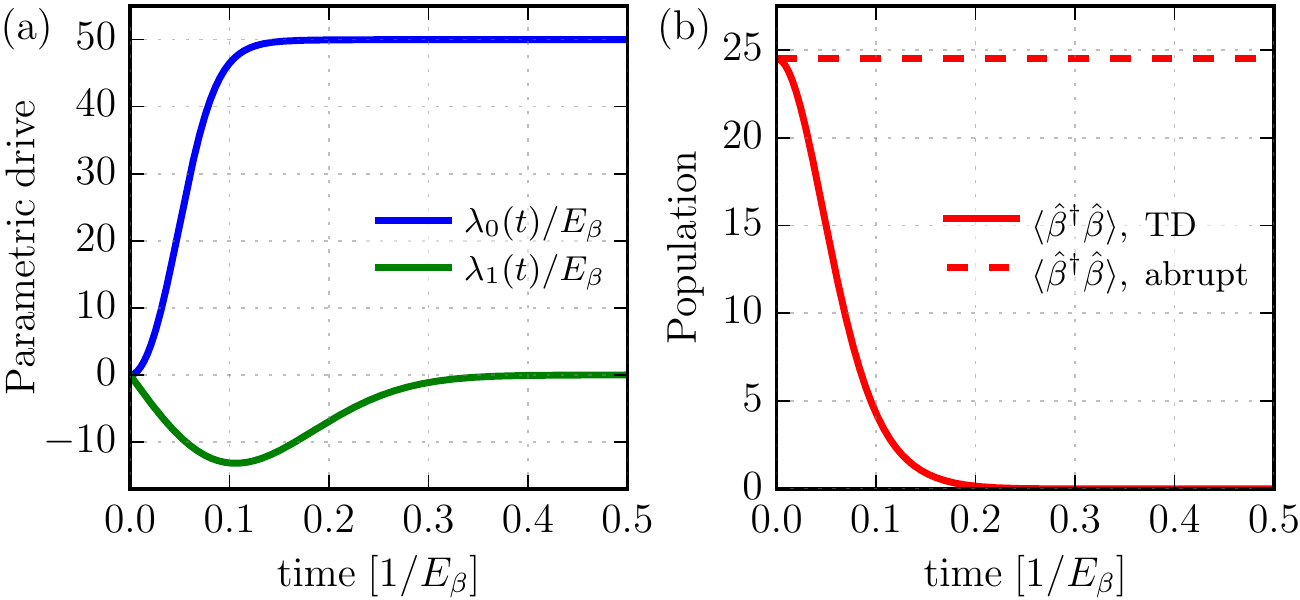}
\caption{\textbf{Initialization of the mechanical state.} 
Fast turn on of mechanical parametric driving using the transitionless driving (TD) scheme (see main text).  
(a) Time dependence of the parametric driving strength $\lambda(t)=\lambda_0(t)+i\lambda_1(t)$, corresponding to a Gaussian profile for the 
instantaneous amplification factor $r(t)$ ($\tanh 2r(t) = \lambda_0 (t)/\Delta$).  The final value of $\lambda(t)$ corresponds to $e^{2r}=20\,$dB.  
The pulse is chosen to ramp up the parametric drive in a time much shorter than the inverse Bogoliubov-mode energy $\Eb$.   
(b) Evolution of the mechanical state, as characterized by the population of the Bogoliubov mode $\hat{\beta}$.
The solid red line is for the TD approach, showing preparation of a pure squeezed state (characterized by no $\beta$-mode excitations) in a time $\sim0.1/\Eb$.
In contrast, a sudden (step-function) turn-on of the parametric drive results in $\beta$ mode being far from its ground state (red dashed line).
The TD protocol plays a crucial role in our scheme, as it allows a rapid-turn on of the mechanically-mediated photon-photon interaction, without any spurious effects resulting from a large initial
$\beta$-mode population.  Neither a purely adiabatic protocol, nor a sudden diabatic approach would be sufficient.  
Here the mechanical dissipation is $\gamma=10^{-4}\Eb$ and $g=0$, but the results are unchanged for $g\neq0$ and a sufficiently small~$\gamma$.
}
\label{Fig:TD}
\end{figure}

\blank
\textbf{Dissipation and mechanical state preparation.}
In addition to the coherent dynamics described by Eq.~\eqref{Eq:Hinitial}, we take into account the coupling of the MR and both optical modes to Markovian baths;
these cause the cavities to be damped at a rate $\kappa$, and the mechanics at a rate $\gamma$.
In the presence of a parametric drive, the noise coming from the MR bath is also amplified.
In the weak mechanical dissipation limit ($\gamma\ll\Eb$), a MR bath of thermal occupancy $\nthM$ corresponds to a bath for the $\beta$ mode of effective temperature $\nthb = \nthM \cosh 2r + \sinh^2r$~\cite{SMnatcom}.
For mechanical excitations off-resonant with the optical modes, i.e.~$\Eb-\Deltaot > \kappa, \tilde{g}$ and $e^{2r}\gg1$, the cavities are heated through the OM interaction at a rate $\Gamma \propto \gamma[\tilde{g}e^r/(\Eb-\delta)]^2(2\nthM+1)$~\cite{SMnatcom}; left unchecked, this heating could corrupt any nonclassical behaviour induced by the enhanced single-photon OM interaction. 
To circumvent amplified noise from the mechanical bath, 
a possible strategy is to add an optical mode to the system, and use it to keep the Bogoliubov mode in its ground state via dissipative squeezing~\cite{Cirac_Zoller_PRL_1993, Rabl_PRB_2004, Parkins_PRL_2006, DallaTorre_PRL_2013, Tan_PRA_2013, Didier_PRA_2014, Kronwald_PRA_2013_DissSqueezing}.
This steady-state technique has recently been implemented experimentally~\cite{Wollman_2015, Pirkkalainen_ArXiv_2015}.

As an alternative to using an additional optical mode, one can instead take advantage of the tunability of the parametric drive.
Indeed, one can first turn on the parametric drive on a timescale $\tau_\mathrm{on}$ short enough to avoid significant perturbation of the initial photon state, i.e.~$\tau_\mathrm{on}<1/\kappa,1/\tilde{g}$.
Then, one can let the system interact for a time $\tau$ sufficient to observe nonclassical signatures, i.e.~$\tau > 1/\tilde{g}$.
This protocol has to be performed in a total time short enough to avoid unwanted cavity heating, i.e.~$\tau_\mathrm{on} + \tau <1/\Gamma$.
This is possible given that $\Gamma$ remains $\ll\tilde{g}$ even for large enhancement factors $e^{2r}\gg1$, as the intrinsic mechanical damping $\gamma$ is extremely low in state-of-the-art experiments.

In such pulsed schemes, it is crucial that 
the initial ramp of the parametric drive amplitude prepares the $\beta$ mode reasonably close to its ground state; this is needed to obtain the radiation pressure interaction in Eq.~\eqref{H2PT} (i.e.~to be able to neglect $\hat{H}'_\mathrm{SRP}$).
If the mechanics $\hat{b}$ remains in its ground state, as it would occur for an abrupt turn on of the parametric drive, then the $\beta$ mode is in a highly squeezed state and this squeezing completely negates the exponential enhancement of the interaction in Eq.~\eqref{H2PT}.
While an adiabatic protocol would prevent $\beta$-mode squeezing, it would be too slow to prevent important perturbation of initial cavity states.
Indeed, for $\Eb \sim \kappa, \tilde{g}$, adiabaticity is ensured for turn on times much longer than $1/\Eb\sim1/\kappa,1/\tilde{g}$.
An appropriate solution is to use the so-called ``counterdiabatic'' or  ``transitionless'' driving (TD) protocols  \cite{Demirplak2003,Demirplak2008,Berry_2009}.  These require one to control the amplitude and the phase of the parametric drive, $\lambda(t) = \lambda_0(t) + i\lambda_1(t)$ [cf.~Fig.~\ref{Fig:TD}~(a)].
The term $\lambda_0(t)$ defines the instantaneous Bogoliubov mode of interest through $\tanh 2r(t) = \lambda_0 (t)/\Delta$, with $\lambda(0) =0$ and $\lambda(\tau_\mathrm{on})=\lambda$.
The correction $\lambda_1(t)= -\dot{r}(t)$ ensures that the MR stays in the ground state of the instantaneous Bogoliubov mode despite non-adiabatic effects.
In Fig.~\ref{Fig:TD}~(b), we show the evolution of the $\beta$ mode without nonlinear interaction ($g=0$) and for a MR initially in its ground state.
Using TD,
the final $\beta$ mode is prepared in its ground state for $\tau_\mathrm{on} \ll 1/\kappa$.
Such an ideal preparation is not possible if one just suddenly turns on $\lambda(t)$.

\blank
\textbf{Standard radiation pressure interaction.}
We focus in the remainder of the paper on the case where the relative detuning 
$\delta = 0$, such that the OM interaction is described by $\hat{H}_\mathrm{SRP}$, Eq.~\eqref{H2PT}; we further take parameters such that $\Eb > \kappa, \tilde{g}$ to ensure a sufficiently wide-bandwidth mechanically-mediated photon-photon interaction.
This two-photon interaction can be understood as an effective ``feedback" process:  the photonic system first displaces the MR and then this displacement 
results in an effective forcing of the photonic system~\cite{Bose_PRA_1997, Rabl_PRL_2011, Nunnenkamp_PRL_2011, Ludwig_Safavi_PRL_2012}. 
The conventional approach 
to describing such an interaction uses a polaron transformation 
$\hat{U}=\exp\{(\tilde{g}/\Eb)(\atd\ao+\aod\at)(\bd-\b)\}$ on $\hat{H}_\mathrm{SRP}$,
leading to the polaron Hamiltonian $\hat{H}_\mathrm{SRP}^\mathrm{pol}=\hat{U}\hat{H}_\mathrm{SRP}\hat{U}^\dag$,
\begin{subequations}
\begin{align}
\hat{H}_\mathrm{SRP}^\mathrm{pol} 
	& = \Eb\bd\b-\Lambda(\atd\ao+\aod\at)^2, \\
	& = \Eb\bd\b-\Lambda(\hat{a}_\mathrm{s}^\dag\hat{a}_\mathrm{s}-\hat{a}_\mathrm{a}^\dag\hat{a}_\mathrm{a})^2,
\label{HPol}
\end{align}
\end{subequations}
with $\Lambda = \tilde{g}^2/\Eb$.
Eq.~\eqref{HPol} is written in the symmetric/antisymmetric photonic basis, defined by the modes
$\hat{a}_{\mathrm{s},\mathrm{a}}=(\hat{a}_1\pm\hat{a}_2)/\sqrt{2}$.
When only the symmetric mode is driven, the nonlinearity is a Kerr interaction and the physics of the radiation pressure interaction in a single-cavity OM system is recovered.
As described in Refs.~\onlinecite{Rabl_PRL_2011, Nunnenkamp_PRL_2011}, the polaron transformation only diagonalises the Hamiltonian of the closed system. 
When including dissipation or a drive, the finite displacement of the $\beta$ mode caused by the photons has to be accounted for 
($\langle \b \rangle = \tilde{g}/\Eb \langle\hat{a}_\mathrm{s}^\dag\hat{a}_\mathrm{s}-\hat{a}_\mathrm{a}^\dag\hat{a}_\mathrm{a}\rangle$).
As a result, when a photon enters or leaves a cavity, it generates phonon sideband excitations (i.e. excitations of the $\beta$ mode); this is analogous to standard Franck Condon physics. 

\begin{figure}[t]
\center
\includegraphics[width=\columnwidth]{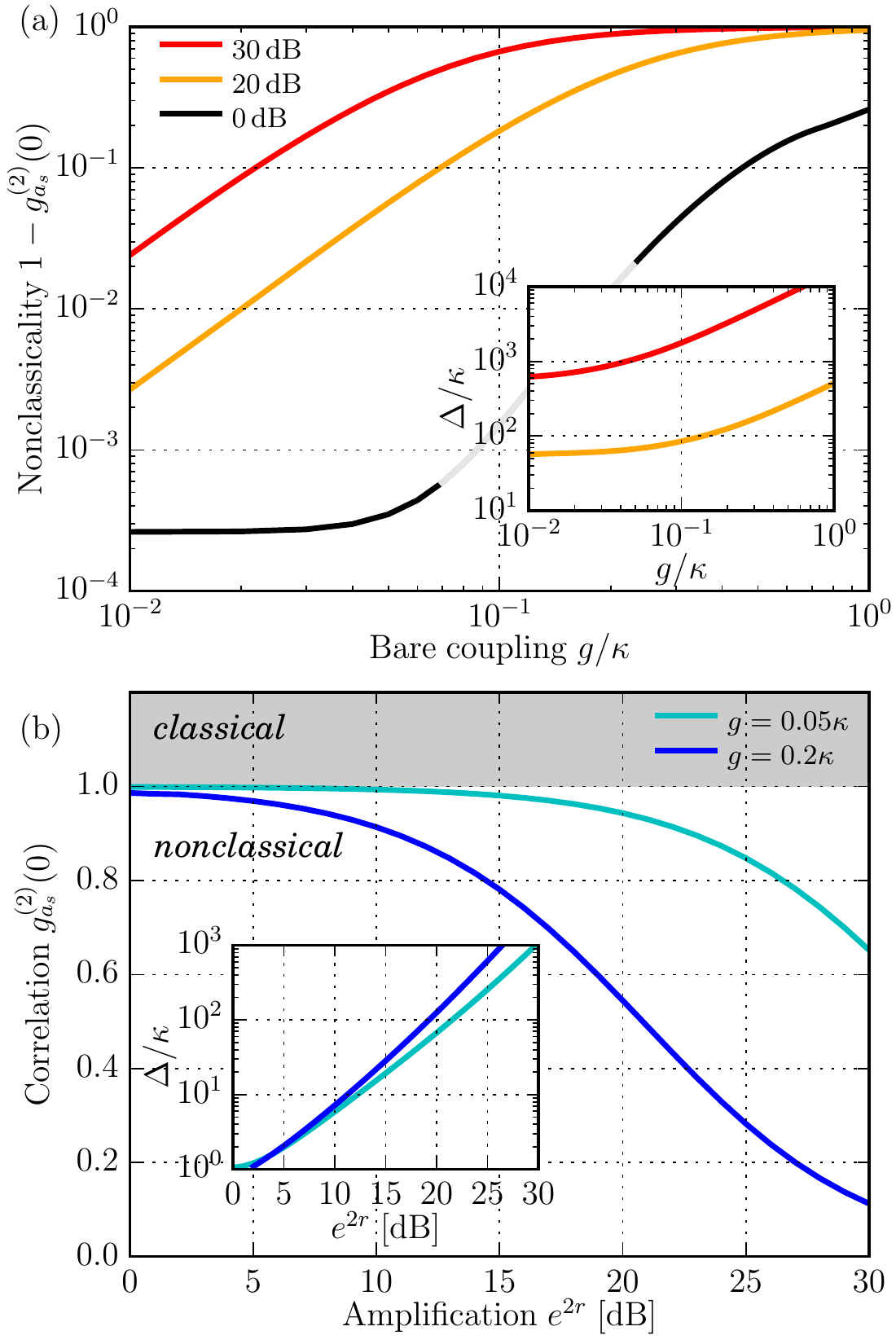}
\caption{\textbf{Nonclassical photon intensity correlations.}
Intensity correlation function $g_{a_s}^{(2)}(0) = \langle \hat{a}_s^\dag\hat{a}_s^\dag\hat{a}_s\hat{a}_s\rangle / \langle \hat{a}_s^\dag\hat{a}_s \rangle^2$, 
for a mechanical parametric drive yielding the optomechanical interaction $\hat{H}_\mathrm{SRP}$ [cf.~Eq.~\eqref{H2PT}], 
and for a weak coherent probe tone applied to the cavities.
In (a), $g_{a_s}^{(2)}(0)$ is plotted as a function of $g$ for different values of 
the amplification factor $e^{2r}$; in (b), it is plotted as a function of $e^{2r}$ for a fixed value of $g$.  
For each value of $g$ and $r$, the probe frequency and the parametric-drive detuning $\Delta$ are optimized to minimize $g_{a_s}^{(2)}(0)$ ($\Delta$ is plotted in insets).
The amplitude of the weak probe is kept fixed.
Panel (a) demonstrates that the violation the classical bound $g_{a_s}^{(2)}(0) \geq 1$ is enhanced in our scheme:  
the presence of the mechanical parametric drive leads to significant suppression of $g_{a_s}^{(2)}(0)$ for experimentally accessible couplings $g \sim 0.01\kappa$.
Panel (b) shows that mechanical parametric driving brings the optical field deep into the nonclassical region even for $g=0.05\kappa$.
For these results, dissipative squeezing is used, 
with a damping rate $\gamma_\beta=0.001\kappa$. 
}
\label{Fig:g2}
\end{figure}

\blank
\textbf{Photon Blockade.}
The photon-photon interaction in Eq.~\eqref{HPol} can lead to photon blockade,
a quantum phenomenon characterized by a strong suppression of the probability of having more than one photon in the cavity together with antibunched photon statistics.
It has been thoroughly studied in the single cavity setup~\cite{Rabl_PRL_2011};
here we highlight the advantage of parametrically driving the MR.
Photon blockade is typically quantified by the equal-time intensity correlation function
$g^{(2)}_a(0) = \langle \hat{a}^\dag\hat{a}^\dag\hat{a}\hat{a}\rangle / \langle \hat{a}^\dag\hat{a} \rangle^2$
that drops below the classical bound, $g^{(2)}_a(0)<1$.
Note that, although $g^{(2)}_a(0)<1$ can be obtained with Gaussian states obeying a linear dynamics~\cite{Lemonde_PRA_2014},
here the $g^{(2)}_a(0)$ suppression cannot be reproduced if the interaction $\hat{H}_{\mathrm{SRP}}$ is linearized~\cite{SMnatcom}.

\begin{figure}[t]
\centering
\begin{tabular}{l}
\includegraphics[width=\columnwidth]{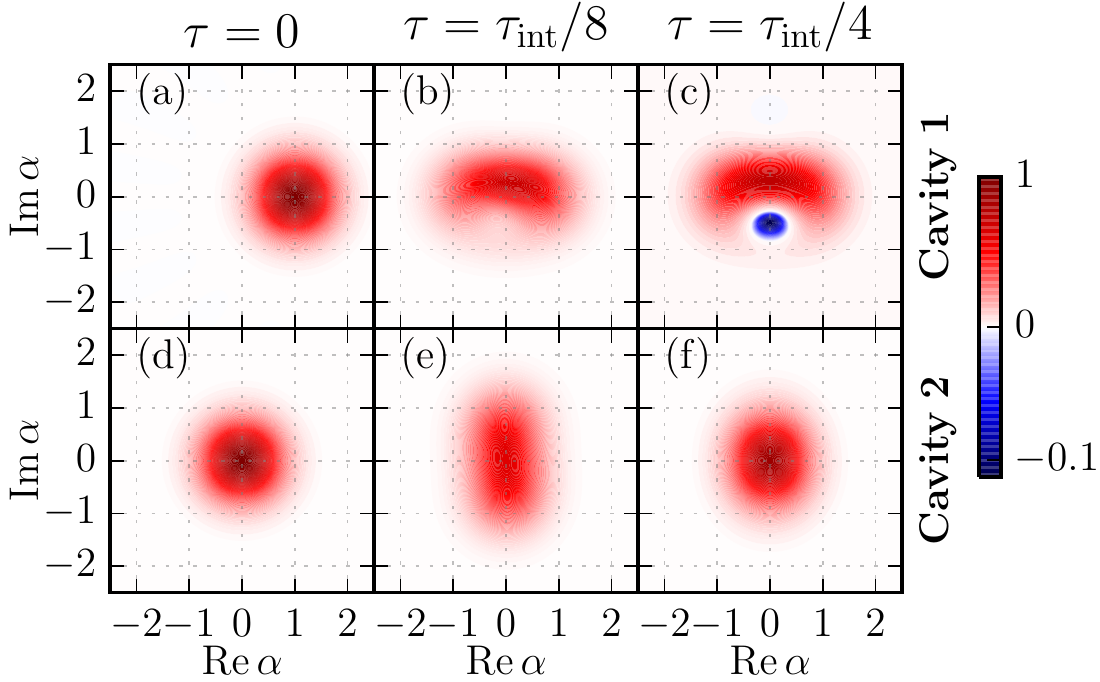}\\
\includegraphics[width=\columnwidth]{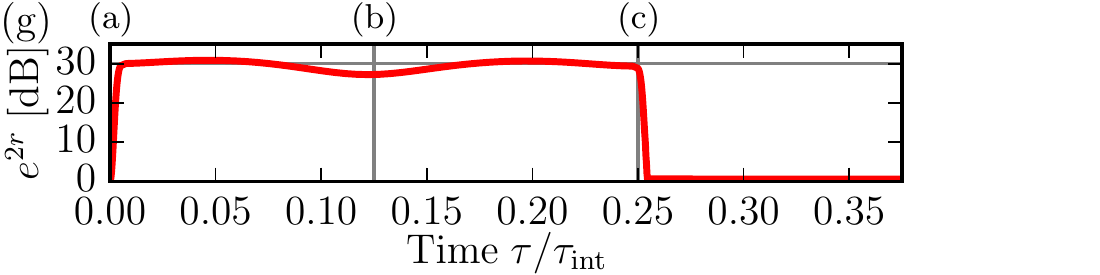}
\end{tabular}
\caption{\textbf{Emergence of negative photonic Wigner functions from enhanced optomechanical coupling}.
Results illustrating the pulsed protocol described in the main text,
for a mechanical parametric drive yielding the optomechanical interaction $\hat{H}_\mathrm{SRP}$
[c.f.~Eq.~\eqref{H2PT}].
The effective photon-photon interaction strength $\Lambda = \tilde{g}^2 / \Eb$ sets the characteristic time $\tau_\mathrm{int} = 2\pi/\Lambda$.
The Wigner function of cavity 1 (2) is plotted in panels (a)-(c) [(d)-(f)] at 3 characteristic times.
Negative (positive) values of the Wigner functions are plotted in blue (red).
(a),(d):  Initial state, where cavity 1 is initially displaced by $\alpha_1=1$, cavity 2 is in vacuum. 
The parametric drive is then switched on using the transitionless driving scheme with a short turn on time 
$\tau_\mathrm{on}=\tau_\mathrm{int}/400$ and a Gaussian profile for $r(t)$, 
the corresponding mechanical amplification strength is plotted in panel (g).
(c) Negativity in the cavity-1 Wigner function is maximal at $\tau=\tau_\mathrm{int}/4$.  As discussed in the main text [and shown in (g)], the parametric drive
can then be turned off with the TD scheme, and the cavity-1 state will be emitted into the cavity-1 input-output waveguide, 
resulting in a propagating photonic state with a negative Wigner function.
Parameters here are $g = 0.3 \kappa$, mechanical damping $\gamma = 10^{-4}\kappa$, and mechanical bath occupancy $\nthM=0.5$;
the parametric drive strength and detuning are chosen to yield an amplification factor $e^{2r} = 30\,\mathrm{dB}$ and $\Eb=2\tilde{g}$.
The resulting Kerr-interaction strength is then $\Lambda \approx 2.4\kappa$ and the rate at which the mechanical noise heats the cavities is $\Gamma \sim \kappa/40$.
If one reduces the amplification factor to 25$\,$dB, the negativity is lost; this highlights the crucial role of the parametric driving.
}
\label{Fig:wigner}
\end{figure}

The intensity correlation of the symmetric mode, $\gts$, is calculated in presence of a weak probe drive on $\hat{a}_s$. 
We use a standard quantum master equation to describe the coherent dynamics governed by $\hat{H}_\mathrm{SRP}$ and the dissipation to zero temperature baths of the $\b$ and $\hat{a}_{1,2}$ modes.
We thus assume that the MR is either cooled using dissipative squeezing, or has been prepared in its ground state via the TD protocol.
The resulting $\gts$, with and without mechanical parametric drive, are compared in Fig.~\ref{Fig:g2}~(a).
The parametric drive dramatically reduces $\gts$, especially in the experimentally accessible regime $g<0.1\kappa$, 
e.g.~for $g=0.1\kappa$, $\gts\approx0.8$ [$\gts\approx 0.3$] for $20\,\mathrm{dB}$ ($30\,\mathrm{dB}$) of amplification while $\gts\approx 0.999$ without parametric drive.
In the limit $\tilde{g} > \kappa$, $\gts$ is minimized for $\Eb=2\tilde{g}$, i.e.~for a parametric detuning $\DeltaM=\frac{1}{2}ge^{3r}$ [cf.~insets of Fig.~\ref{Fig:g2}].
For $20\,\mathrm{dB}$ of amplification and $g\sim0.1\kappa$, this implies $\Delta\sim100\kappa$.
The optimal $\Eb$ corresponds to the situation where, in the polaron picture [cf.~Eqs.~\eqref{HPol}], the state with 2 symmetric photons ($\vert 2, 0, 0 \rangle$) and the corresponding first phonon sideband ($\vert 2, 0, 1 \rangle = \b^\dag\vert 2, 0, 0 \rangle$) are equally detuned from the one-photon state ($\vert 1,0,0 \rangle$).
The intensity correlations are described to a good approximation by $\gts\approx1/[1+0.8(\tilde{g}/\kappa)^2]$.
Increasing the parametric drive is thus, in principle, always beneficial since, for any coupling $g<\kappa$, there is always an amplification strength $r$ that leads to a desired $\gts<1$.
For instance, $\gts=0.5$ is obtained for $e^r\approx2.2\kappa/g$.

\blank
\textbf{Negative Wigner functions.}
The possibility for time-dependent control of the photon-photon interaction in
our system opens the door to a wealth of interesting functionalities. 
Perhaps the most demanding challenge is the production of photon states exhibiting strongly negative Wigner functions.
We show here how this can be accomplished in our setup, in a manner that produces negativity both in the states of intracavity and propagating photons.  Crucially, this can be done using a bare coupling $g$ that is still smaller than the cavity damping rate ($g/\kappa \sim 0.3$).  We stress that this kind of negative Wigner function generation would be essentially impossible without mechanical parametric driving:  not only would one require a $g$ that is at least an order of magnitude larger, one would need some alternate means for controlling it in time.  Our scheme thus significantly lowers the level of experimental improvement needed for generating negative photonic Wigner functions.

One first prepares cavity 1 in a low-amplitude 
coherent state using a classical laser drive while cavity 2 remains in vacuum [Fig.~\ref{Fig:wigner}~(a)].   
The mechanical parametric drive is off during this step, so that there is essentially no photonic nonlinearity.  Once this initial cavity state is prepared, the cavity drive is turned off, and the photonic interaction is amplified by ramping up the mechanical parametric drive. The TD scheme described earlier allows this turn-on step to be completed in a time $\tau_\mathrm{on} \ll 1/\kappa, 1/\tilde{g}$, i.e.~fast enough to be effectively instantaneous to the photons.  At the same time, this scheme ensures that the $\beta$ mode is prepared in its ground state.

We pick a frequency detuning of the parametric drive $\delta=0$ to realize the two-photon tunneling interaction $\hat{H}_\mathrm{SRP}$ [c.f.~Eq.~(\ref{H2PT})].
The effective Hamiltonian $\hat{H}_\mathrm{SRP}^\mathrm{pol}$ leads (in the absence of dissipation) to a periodic evolution with the characteristic time $\tau_\mathrm{int}=2\pi/\Lambda$,
with $\Lambda=\tilde{g}^2/E_\beta$.  If $\Lambda\gg\kappa$ (possible with large enough parametric driving), one finds that the cavity-1 state is strongly nonclassical at $\tau\sim \tau_\mathrm{int}/4$, characterized by a Wigner function exhibiting large amounts of negativity [Fig.~\ref{Fig:wigner}~(c)].
This can be easily understood by considering the effects of the two-photon tunneling interaction that is mediated by the mechanics, 
$\hat{a}_2^{\dag} \hat{a}_2^{\dag}\hat{a}_1 \hat{a}_1 + \mathrm{H.c.}$  As cavity 2 starts in vacuum and cavity 1 has negligible probability for having more than 2 photons,  this term initially transfers two photons from the first to the second cavity in a time $\tau_\mathrm{int}/8$.  The two-photon Fock state of cavity 2 then gets weakly populated and its Wigner function is reminiscent of a low-amplitude squeezed state [Fig.~\ref{Fig:wigner}~(e)].  After an additional evolution for a time $\tau_\mathrm{int}/8$, these two photons return to cavity 1, with an overall $\pi$-phase shift.  This phase shift of the 2-photon component of the cavity-1 state (with respect to the 1 photon component) leads to negativity in the Wigner function [Fig.~\ref{Fig:wigner}~(c)].  
 
Next, at the special time $\tau_\mathrm{int}/4$ where the cavity-1 state is maximally nonclassical, the parametric drive is rapidly turned off.  By using the reverse of our TD protocol
 [cf.~Fig.~\ref{Fig:wigner}~(g)], this can be done in such a way that the MR returns to its ground state.  At this stage, the nonlinear optomechanical interaction is almost completely suppressed:
 not only is its magnitude greatly diminished, but it is now no longer resonant, such that any residual effects will scale as $g^2 / \omega_M \ll \kappa$ (see \cite{SMnatcom} for more details).  Finally, in the ideal case where internal cavity losses are weak, the nonclassical cavity-1 state is converted perfectly to a propagating mode in the cavity-1 input-output waveguide with an exponential profile. 
We thus have generated a nonclassical, propagating photonic state, using an underlying weak single-photon optomechanical coupling and the additional linear resource of a parametric drive.
We stress that the ability to rapidly turn the mechanically-mediated nonlinear interaction on and off is crucial to being able to do this experiment.

\blank
\textbf{Engineered MR response function.}
While our treatment so far is rigorous, the origin of the enhanced OM interaction may still seem somewhat mysterious.  An alternate approach which provides a more intuitive picture, and which is more easily generalized to more complex systems, is based on deriving an effective Keldysh action for the cavity photons.  In this approach,
one clearly sees that 
the mechanical resonator mediates a time-nonlocal effective photon-photon interaction, that depends crucially on the retarded Green's functions of the mechanics.

Indeed, by 
integrating out the mechanical degree of freedom in the Keldysh action obtained for the interaction $\hat{H}_\mathrm{SRP}$, 
one gets an effective action describing two distinct time-non-local photon-photon interactions.
These interactions can equivalently be captured by writing the equations of motion for the cavity fields; for cavity 1, the interaction term in the equation of motion is:
\begin{align}
\dot{\hat{a}}_1(t)=\int_{-\infty}^{+\infty}\!\mathrm{d}t'2i\{&\Lambda(t-t')\hat{a}_2^\dag(t')\hat{a}_1(t')\hat{a}_2(t)\nonumber\\
+&\tilde{\Lambda}(t-t')\hat{a}_1^\dag(t')\hat{a}_2(t')\hat{a}_2(t)\}+\dots
\label{EqEOMeff}
\end{align}
with 
$\Lambda(t) = \frac{1}{2}g^2G^R_b(t)$,
$\tilde{\Lambda}(t) = \frac{1}{2}g^2\tilde{G}^R_b(t)$. 
Here, 
$G^R_b(t) = -i\theta(t) \langle[ \hat{b}(t) , \hat{b}^\dag (0) ]\rangle$  and 
$\tilde{G}^R_b(t) = -i\theta(t)\langle [ \hat{b}(t) , \hat{b} (0) ]\rangle$ 
are respectively the non-interacting 
diagonal and off-diagonal retarded mechanical Green's functions.
The role of the parametric drive is to render the off-diagonal element non-zero and amplify $G^R_b(t)$ and $\tilde{G}^R_b(t)$.
For large amplification, 
$G^R_b(t)=\tilde{G}^R_b(t)$
with, in the Fourier domain,
\begin{equation}
\tilde{G}^R_b[\omega]=\frac{e^{2r}}{4}\left[\frac{1}{\omega - \Eb + i\gamma/2} - \frac{1}{\omega + \Eb + i\gamma/2} \right].
\end{equation}
In the limit where $\Eb \gg \kappa, \tilde{g}$, the frequency dependence of the interaction is not important on the relevant energy scales of the system and can be neglected. In this case, the situation is similar to an instantaneous interaction and one recovers the polaron picture of Eq.~\eqref{HPol}.
This description 
clearly shows the general idea: 
to amplify the effective photon-photon interaction, one has to engineer the dynamics, i.e.~the response function, of the MR.

\blank
\textbf{Conclusions.}
We have studied a two-cavity OM system, showing that parametrically driving the MR exponentially enhances the nonlinear OM interaction. 
One can thus reach the much-coveted single-photon strong coupling regime starting from an extremely weak bare interaction $g$.  This allows 
photon blockade and non-Gaussian state generation even when $g \ll \kappa$.
This new scheme further benefits from its controllability: one can choose the nature of the nonlinear interaction to amplify as well as modulate in time its strength.
Our work suggests more general approaches for enhancing bosonic-mediated interactions and nonlinearities through simple parametric driving.

\blank
\textbf{Acknowledgements.}
This work was supported by NSERC.

\blank
\textbf{Methods}\\
\small{
\textbf{Transitionless driving.}
We give here more details about the ``counterdiabatic'' or ``transitionless'' driving protocols  \cite{Demirplak2003,Demirplak2008,Berry_2009}.
These imply controlling the amplitude and the phase of the parametric drive, such that $\lambda(t) = \lambda_0(t) + i \lambda_1(t)$, with $\lambda(0) = 0$ and $\lambda(\tau_\mathrm{on}) = \lambda$. 
Defining the instantaneous unitary transformation $\hat{U}(t) = \exp[\frac{1}{2}r(t)(\hat{b}^2-\hat{b}^{\dag 2})]$ with $\tanh 2r(t) = \lambda_0(t)/\Delta$ and considering only the parametrically driven MR, i.e.~$\hat{H}$ given by Eq.~\eqref{Eq:Hinitial} with $g=0$, the transformed Hamiltonian is,
\begin{subequations}
\begin{align}
	\hat{\tilde{H}}(t) & = \hat{U}(t) \hat{H} \hat{U}^\dag(t) + i \dot{\hat{U}}(t)\hat{U}^\dag(t) \\
	& = \frac{\Delta}{\cosh 2r(t)}\hat{b}^\dag\hat{b} + i\tfrac{1}{2}[\lambda_1(t)+\dot{r}(t)](\hat{b}^{\dag 2}-\hat{b}^2).
\end{align}
\end{subequations}
Consequently, starting from the mechanical ground state, a parametric drive modulated with $\lambda_1(t) = -\dot{r}(t)$
ensures that (without dissipation) the instantaneous Bogoliubov mode $\b(t) = \cosh r(t)\hat{b} - \sinh r(t)\hat{b}^\dag$ stays in its ground state. 
At the end of the protocol, the desired $\beta$ mode is thus in a vacuum state.
Considering dissipation, we show in the main text that it is still possible to prepare the final Bogoliubov mode $\b$ in the same state as the initial MR state in a time $\tau_\mathrm{on}$ much faster than any other timescales of the system (see Fig.~\ref{Fig:TD}).

\blank
\textbf{Quantum master equation.}
To obtain the $g^{(2)}_a$ correlation function and the Wigner function of the cavities state, we use a standard quantum master equation approach~\cite{GardinerZollerBook}.
The coherent dynamics is governed by $\hat{H}_\mathrm{SRP}$ [cf.~Eq.~\eqref{H2PT}] and the coupling of the cavities to zero-temperature baths is described with the Lindbladians
$\hat{L}=\kappa D[\hat{a}_1]+\kappa D[\hat{a}_2]$ where $D[\hat{a}]\cdot=\hat{a}\cdot\hat{a}^\dag-\frac{1}{2}\{\hat{a}^\dag\hat{a},\cdot\}$.
Concerning the mechanical Lindbladian, we consider two configurations: either the Bogoliubov mode is cooled down to its ground state with dissipative squeezing or a TD scheme is used with a MR initially in a thermal state (population $\nthM$).
The dissipative squeezing protocol is modelled with the Lindbladian $\gamma_\beta D[\hat{\beta}]$, 
where $\gamma_\beta$ is the coupling rate to the engineered reservoir that squeezes the mechanics, and is used to obtain the results presented in Fig.~\ref{Fig:g2}. 
For these results, we consider a drive on the cavities,
$\hat{H}_\mathrm{drive}=\epsilon(\hat{a}_1e^{i\omega_{d1}t}+\hat{a}_2e^{i\omega_{d2}t}) +\mathrm{H.c.}$; the drive is used to probe the intensity correlations.
Meanwhile, the transitionless driving scheme is used in the protocol that leads to negative Wigner functions of the optical mode [cf.~Fig.~\ref{Fig:wigner}] and corresponds to a Lindbladian 
$\gamma (1+\nthM) D[\hat{b}] + \gamma \nthM D[\hat{b}^\dag]$.
The parametric drive strength is turned on continuously, with $\lambda(t)$ derived above, and we consider an initial coherent state in cavity 1.
The density matrix $\hat{\rho}$ in these two situations is then obtained from the quantum master equation
\begin{equation}
\dot{\hat{\rho}}=-i[\hat{H},\hat{\rho}]+\hat{L}\hat{\rho}.
\end{equation}
The intensity correlations $g_a^{(2)}(0)$ are calculated from the steady state value of $\hat{\rho}$.
For given values of the coupling $g$ and amplification strength $e^{2r}$, the detuning $\Delta$ and the drive frequency are optimized to minimize the intensity correlations.
The results are plotted in Fig.~\ref{Fig:g2}.

\blank
\textbf{Effective Keldysh action.}
As explained in the main text, the finite-time response function of the MR results in a time-nonlocal photon-photon interaction. 
To describe this physics, we calculate the action of the
system in the same interaction picture used for Eqs.~\eqref{H2PT}-\eqref{HPAT}, i.e.~in a frame where the Hamiltonian is not explicitly time-dependent.
Since the OM system is driven and subject to dissipation, the Keldysh formalism is well adapted to study this out-of-equilibrium system~\cite{KamenevBook}; a detailed example of the Keldysh formalism in OM systems is presented in Ref.~\cite{Lemonde_PRA_2015}.
In this approach, each annihilation operator used in the Hamiltonian-based description is mapped onto two time-dependent fields: a classical ($\mathrm{cl}$) and a quantum ($\mathrm{q}$) field.

For the the cavities ($\sigma = 1,2$) and MR fields,
\begin{subequations}
\begin{align}
	\mathbf{a}_\sigma^\dag & \equiv \left(  a^*_{\sigma, \mathrm{cl}}(t) , a_{\sigma, \mathrm{cl}}(t), a^*_{\sigma, \mathrm{q}}(t), a_{\sigma, \mathrm{q}}(t) \right), \\
	\mathbf{b}^\dag & \equiv \left(  b^*_\mathrm{cl}(t) , b_\mathrm{cl}(t), b^*_\mathrm{q}(t), b_\mathrm{q}(t) \right),
\end{align}
\end{subequations}
the Keldysh action that describes the full OM system studied in the main text [cf.~Eq.~(1)] has the general following form,
\begin{align}
S_{\rm{tot}} 
= &\sum_{\sigma = 1,2} \mathbf{a}_{\sigma}^\dag\check{\mathbf{G}}_{a_\sigma}^{-1}(t-t')\mathbf{a}_{\sigma}(t') 
+ \mathbf{b}^\dag\check{\mathbf{G}}_{b}^{-1}(t-t')\mathbf{b}(t') \nonumber\\
+ \frac{g}{\sqrt{2}} &\sum_{ijk = \mathrm{cl,q}}\zeta_{ijk}\int_{-\infty}^{\infty} \mathrm{d}t \left[a^*_{2,i}(t)a_{1,j}(t)b_k(t) + \mathrm{c.c.} \right]. \label{Eq:Stot}
\end{align}
Here, $\zeta_{ijk} = 1$ if there is an odd number of quantum fields and $0$ otherwise.

In Eq.~\eqref{Eq:Stot}, the two first terms represent the Gaussian action that governs the non-interacting dynamics (i.e.~$g=0$). It involves the non-interacting cavities (MR) Green's functions $\check{\mathbf{G}}_{a_\sigma}(t-t')$ [$\check{\mathbf{G}}_{b}(t-t')$].
Here, the most general Green's functions are $4\times4$ matrices of the form
\begin{subequations}
\begin{align}
	\check{\mathbf{G}}_b(t) &= \begin{pmatrix}
		\mathbf{G}_b^K(t) & \mathbf{G}_b^R(t) \\
		\mathbf{G}_b^A(t) & 0
	\end{pmatrix},\\
	\mathbf{G}_b^R(t) &= [\mathbf{G}_b^A(-t)]^\intercal = \begin{pmatrix}
	G^R_b(t) & \tilde{G}^R_b(t) \\
	[\tilde{G}^R_b(t)]^* & [G^R_b(t)]^*
	\end{pmatrix},\\
	\mathbf{G}^K_b(t) &= \begin{pmatrix}
	G^K_b(t) & \tilde{G}^K_b(t) \\
	-[\tilde{G}^K_b(-t)]^* & G^K_b(-t)
	\end{pmatrix}.
\end{align}
\end{subequations}
The retarded Green's functions encode information on the single-particle density of states, and also describe linear response of the system to external perturbations:
\begin{subequations}
\begin{align}
	G^R_b(t) & = -i\theta(t)\langle [ \hat{b}(t), \hat{b}^\dag(0) ] \rangle, \\
	\tilde{G}^R_b(t) & = -i\theta(t)\langle [ \hat{b}(t), \hat{b}(0) ] \rangle.
\end{align} \label{Eq:GR}%
\end{subequations}
The Keldysh Green functions encode information on the distribution functions:
\begin{subequations}
\begin{align}
	G^K_b(t) & = -i\langle \lbrace \hat{b}(t), \hat{b}^\dag(0) \rbrace \rangle, \\
	\tilde{G}^K_b(t) & = -i\langle \lbrace \hat{b}(t), \hat{b}(0) \rbrace \rangle.
\end{align}
\label{Eq:GK}%
\end{subequations}
\indent As the action of Eq.~\eqref{Eq:Stot} only has linear and quadratic terms in the mechanical fields, the MR can exactly be integrated out \cite{KamenevBook}.  The resulting action which describes only photonic degrees of freedom is:
\begin{subequations}
\begin{align}
&S_\mathrm{eff} = S_{a_1}^0 + S_{a_2}^0 +\frac{g^2}{8}\iint_{\mathbb{R}^2}\mathrm{d}t\mathrm{d}t'[s^R(t,t') + s^K(t,t')],\label{Eq:Sfinala}\\
&s^{R,K}(t,t') = \sum_{ijkl=\mathrm{cl,q}} \zeta^{R,K}_{ijkl} \left[ s^{R,K}_{ijkl}(t,t') + \mathrm{c.c.} \right],\label{Eq:Sfinalb}\\
&s^{R,K}_{ijkl}(t,t') = a^*_{2,i}(t)a_{1,j}(t)G_b^{R,K}(t-t')a_{2,k}(t')a^*_{1,l}(t')  \nonumber\\
&\hspace{14.3mm}+ a^*_{2,i}(t)a_{1,j}(t)\tilde{G}_b^{R,K}(t-t')a^*_{2,k}(t')a_{1,k}(t').
\label{Eq:Sfinalc}%
\end{align}
\end{subequations}
Here $\zeta^R_{ijkl} = 1$ if the interaction term has an odd number of quantum fields and $\zeta^R_{ijkl} = 0$ otherwise, while $\zeta^K_{ijkl} = 1$ if there is both one quantum field between the $i,j$ components and one quantum field between the $k,l$ component, i.e.~a total of two quantum fields, and $\zeta^K_{ijkl} = 0$ otherwise. 
The first two terms of Eq.~\eqref{Eq:Sfinala} represent the non-interacting cavities, the third term describes the coherent time-nonlocal photon-photon interaction while the fourth term describes the extra noise that perturbs the cavities due to their interaction with the MR. As one can see, the diagonal (off-diagonal) MR Green function $G^R_b(t)$ [$\tilde{G}^R_b(t)$] mediates a cross Kerr type interaction (two-photon tunneling) between the cavities. From this effective action, it is clear that modifying the MR Green's functions leads to a modification of the effective photon-photon interaction. 

Finally, following Ref.~\cite{KamenevBook}, one can show that the interaction term $s^R$ in the action of Eq.~\eqref{Eq:Sfinala} is equivalent, in the cavity effective equation of motion, to the contribution highlighted in Eq.~\eqref{EqEOMeff}. 
A less elegant alternative approach to obtain this effective equation of motion is to first solve the Heisenberg-Langevin equation for $\b$. 
This solution is used to eliminate the $\b$ from the cavities Heisenberg-Langevin equations.
The effective photon-photon interaction as well as the additional nonlinear noise term then explicitly appear.

\end{document}


\title{Supplementary Information for\\``Enhanced nonlinear interactions in quantum optomechanics\\via mechanical amplification''}

\author{Marc-Antoine Lemonde}
\affiliation{Department of Physics, McGill University, 3600 rue University, Montreal, Quebec H3A 2T8, Canada}
\author{Nicolas Didier}
\affiliation{Department of Physics, McGill University, 3600 rue University, Montreal, Quebec H3A 2T8, Canada}
\affiliation{D\'epartment de Physique, Universit\'e de Sherbrooke, 2500 boulevard de l'Universit\'e, Sherbrooke, Qu\'ebec J1K 2R1, Canada}
\author{Aashish~A.~Clerk}
\affiliation{Department of Physics, McGill University, 3600 rue University, Montreal, Quebec H3A 2T8, Canada}

\maketitle

\renewcommand{\theequation}{S\arabic{equation}}
\renewcommand{\thefigure}{S\arabic{figure}}
\renewcommand{\bibnumfmt}[1]{[S#1]}
\renewcommand{\citenumfont}[1]{S#1}

\section{Derivation of the optomechanical Hamiltonian}

We start by describing in more details the rotating wave approximation performed on the system Hamiltonian leading to Eq.~(1) of the main text. 
The starting Hamiltonian that describes the coherent dynamics of the undriven two-cavity optomechanical (OM) setup of interest has the following form,
\begin{align}
	\hat{H}= \Delta \hat{b}^\dag\hat{b} - \frac{1}{2}(\lambda \hat{b}^2 + \lambda^* \hat{b}^{\dag2}) + g(\hat{a}_2^\dag \hat{a}_1 e^{i(\omega_2 - \omega_1)t} + \hat{a}_1^\dag \hat{a}_2 e^{-i(\omega_2 - \omega_1)t})(\hat{b}e^{-i\omega_pt} + \hat{b}^\dag e^{i\omega_pt}).
\end{align}
$\hat{H}$ is expressed in an interaction picture with respect to the free cavity Hamiltonians ($\omega_1, \omega_2$) and, for the mechanics, with respect to the pump frequency $\omega_p$.
The best way to show the validity of the rotating wave approximation performed in our work is to first express $\hat{H}$ in terms of the eigenmode of the quadratic part, i.e.~the Bogoliubov mode $\b = \hat{b}\cosh r - \hat{b}^\dag\sinh r$. In the interaction picture with respect to the free $\beta$ mode, it reads
\begin{subequations}
\begin{align}
	\hat{H} 
	 = g & \left[ \hat{a}_2^\dag \hat{a}_1 \left( \b\cosh r e^{-i\Eb t} + \bd\sinh r e^{i\Eb t} \right)e^{-i\delta t} + \textrm{H.c.}\right] \label{Eq:HRWAa} \\
	+ g & \left[ \hat{a}_2^\dag \hat{a}_1\left( \bd\cosh r e^{i\Eb t} + \b\sinh r e^{-i\Eb t} \right)e^{i(\omega_p + \omega_{21}) t} + \textrm{H.c.}\right], \label{Eq:HRWAb}%
\end{align}
\label{Eq:HRWA}%
\end{subequations}
with $\omega_{21} = \omega_2 - \omega_1$, $\delta = \omega_p - \omega_{21}$ and $\Eb = \Delta/\cosh 2r$.
As a consequence, for $\omega_p + \omega_{21} \gg \Eb, \tilde{g}, \delta, \kappa, \gamma$, where $\gamma$ and $\kappa$ are the mechanical resonator (MR) and the cavities damping rates respectively and $\tilde{g} = g e^r /2$, the terms oscillating at $\omega_{21} + \omega_p \pm \Eb$ [Eq.~\eqref{Eq:HRWAb}] can be safely neglected compared to the terms oscillating at $\delta \pm \Eb$.
In that case, the resulting Hamiltonian,
\begin{equation}
	\hat{H} \approx g\hat{a}_2^\dag \hat{a}_1 (\b\cosh r e^{-i\Eb t} + \bd\sinh r e^{i\Eb t})e^{-i\delta t} + \textrm{H.c.},
	\label{HRWA2}
\end{equation}
is exactly the Hamiltonian of Eq.~(1) of the main text, expressed in the interaction picture with respect to the free $\beta$ mode.

In our scheme, we are particularly interested in large amplifications $r$ and for detunings such that $\delta = 0$ ($\omega_{21} = \omega_p$).
The optimal parameters which then lead to the most pronounced quantum signatures imply $\Delta \gg \Eb, \tilde{g}, \kappa, \gamma$ and $\Eb \sim \kappa, \tilde{g}$.
It is thus possible to choose $\omega_p + \omega_{21} = 2\omega_p$ big enough so  that the rotating wave approximation is always valid. For a positive detuning $\Delta >0$, as we have considered all along this work, it constrains $\omega_M - \Delta \gg \kappa, \tilde{g}, \gamma$. 
Choosing negative detunings $\Delta < 0$ relaxes this constraint but might cause other issues, like the possibility to excite additional mechanical mode with the parametric drive.

In the situation where the parametric drive is turned off ($r=0$), for instance to freeze the dynamics once a negative Wigner function function is obtained (see main text or Sec.~\ref{SecWigner} for details), $\Eb \rightarrow \Delta \gg \kappa, g, \gamma$. 
In this case, both interaction terms of Eqs.~\eqref{Eq:HRWA} become off resonant:
The term Eq.~\eqref{Eq:HRWAa} oscillates at frequency $\Delta$ and
the one of Eq.~\eqref{Eq:HRWAb} at the larger frequency $\Delta + 2\omega_p$.
The contribution from Eq.~\eqref{Eq:HRWAb} can thus be safely neglected for any parametric drive strength.

\section{Quantum dynamics of the parametrically driven mechanics}

\subsection{Langevin equation}

We now derive and solve the Heisenberg-Langevin equations of motion for the MR parametrically driven at frequency $2\omega_p$. At this point, we do not include the optomechanical interaction, i.e.~$g=0$ in Eq.~(1) of the main text. Using standard input-output formalism~\cite{Clerk_RMP_2010} to include dissipation to a Markovian bath of thermal occupancy $\nthM$ and damping rate $\gamma$, we get, in a frame rotating at $\omega_p$,
\begin{align}
	\begin{pmatrix}\hat{b}[\omega] \\ \hat{b}^\dag[\omega] \end{pmatrix}
	= 
	\begin{pmatrix} 
	i(\Delta - \omega) + \frac{\gamma}{2} & -i\lambda \\
	i \lambda & -i(\Delta + \omega) + \frac{\gamma}{2}
	\end{pmatrix}^{-1}
	\begin{pmatrix} \sqrt{\gamma}\hat{\eta}[\omega] \\ \sqrt{\gamma}\hat{\eta}^\dag[\omega] \end{pmatrix}
	\equiv \chi[\omega] \begin{pmatrix} \sqrt{\gamma}\hat{\eta}[\omega] \\ \sqrt{\gamma}\hat{\eta}^\dag[\omega] \end{pmatrix}. \label{Eq:EOM}
\end{align}
Here, $\hat{\eta}$ corresponds to the annihilation operator of the noise coming from the MR dissipative bath. As is standard, we consider white Gaussian noise with zero mean ($\langle \hat{\eta} \rangle = 0$), so that the non-zero correlators are
\begin{equation}
	\langle \hat{\eta}^\dag[\omega] \hat{\eta}[\omega'] \rangle = 2\pi\nthM\delta(\omega+\omega'), \qquad \langle \hat{\eta}[\omega] \hat{\eta}^\dag[\omega'] \rangle = 2\pi(\nthM + 1)\delta(\omega+\omega'). \label{Eq:bathCorr}
\end{equation}

We now briefly present how the equations of motion~\eqref{Eq:EOM} translate in terms of the Bogoliubov mode $\b$ and discuss the consequences on the dissipation of the $\beta$ mode.
The Bogoliubov mode is expressed in terms of the noise operator $\hat{\eta}_\beta[\omega] = \eta[\omega] \cosh r - \eta^\dag[\omega] \sinh r$ as follows:
\begin{align}
	\begin{pmatrix}\b[\omega] \\ \bd[\omega] \end{pmatrix}
	= 
	\textbf{M}
	\begin{pmatrix}\hat{b}[\omega] \\ \hat{b}^\dag[\omega] \end{pmatrix}
	=
	\chi_\beta[\omega]
	\begin{pmatrix} \sqrt{\gamma}\hat{\eta}_\beta[\omega] \\ \sqrt{\gamma}\hat{\eta}_\beta^\dag[\omega] \end{pmatrix}, \label{Eq:EOMbeta}
\end{align}
with
\begin{align}
	\textbf{M} &= \begin{pmatrix}\cosh r & -\sinh r \\ -\sinh r & \cosh r \end{pmatrix},&
	\chi_\beta[\omega] &= \textbf{M}\chi[\omega] \textbf{M}^{-1} = 
	\begin{pmatrix} 
	i(\Eb - \omega) + \frac{\gamma}{2} & 0 \\
	0 & -i(\Eb + \omega) + \frac{\gamma}{2}
	\end{pmatrix}^{-1}.
\end{align}
The equation of motion Eq.~\eqref{Eq:EOMbeta} shows that when the MR ($\hat{b}$) is coupled to a Markovian bath of thermal occupancy $\nthM$, the $\beta$ mode is driven by a squeezed reservoir of thermal population $\nthM$.
However, as this squeezing is far from the $\beta$-mode resonance $\Eb$ (it is at zero frequency in the current rotating frame), it effectively looks like thermal noise with thermal occupation $\nthb = \nthM \cosh 2r + \sinh^2 r$. 
This is precisely revealed in the instantaneous covariance matrix of the $\beta$ mode,
\begin{subequations}
\begin{align}
	\langle \bd\b \rangle (t) 	& = \gamma\iint\frac{d\omega}{2\pi}\frac{d\omega'}{2\pi}\frac{\langle \hat{\eta}_\beta^\dag[\omega']\hat{\eta}_\beta[\omega] \rangle e^{-i(\omega+\omega')t}}{(i\omega'+i\Eb-\gamma/2)(i\omega-i\Eb-\gamma/2)} 
												 = \nthM \cosh 2r + \sinh^2 r, \\
	\langle \b\b \rangle (t) 	& = \gamma\iint\frac{d\omega}{2\pi}\frac{d\omega'}{2\pi}\frac{\langle \hat{\eta}_\beta[\omega']\hat{\eta}_\beta[\omega] \rangle e^{-i(\omega+\omega')t} }{(i\omega'-i\Eb-\gamma/2)(i\omega-i\Eb-\gamma/2)}
												 = \frac{-i\gamma(\nthM+\frac{1}{2}) \sinh 2r}{2\Eb - i\gamma} 
\label{Eq:Meanbb}
\end{align}
\end{subequations}
Here, we have used the correlation functions of the MR bath given in Eqs.~\eqref{Eq:bathCorr}. The fact that the squeezing of the reservoir is off-resonant with the $\beta$ mode explicitly appears in the pre-factor $\gamma/\Eb$ in Eq.~\eqref{Eq:Meanbb}. In the limit of weak dissipation, i.e.~$\gamma/\Eb \ll 1/\sinh 2r$, the $\beta$ mode can thus be considered as being coupled with a damping rate $\gamma$ to a Markovian thermal bath of effective temperature $\nthb$.

\subsection{Estimation of the cavity heating rate}

By adding the coupling to the cavities, one can estimate the rate at which the cavities are heated by the amplified mechanical noise $\hat{\eta}_{\b}(t)$. For simplicity, we consider the particular case where the dynamics is well described by $\hat{H}_\mathrm{SRP}$ [Eq.~(2) of the main text] with $\hat{H}'_\mathrm{SRP} = 0$, i.e.~for $\delta = 0$, $e^{2r} \gg 1$ and for a $\beta$ mode initially close to its ground state. In this case, the equations of motion are given by: 
\begin{subequations}
\begin{gather}
	\dot{\b}  = -(i\Eb + \tfrac{1}{2}\gamma)\b - i\tilde{g}( \atd\ao + \aod\at) -\sqrt{\gamma}\hat{\eta}_{\b}, \\
	\dot{\hat{a}}_{1,2} = -\tfrac{1}{2}\kappa\hat{a}_{1,2} - i\tilde{g}\hat{a}_{2,1}(\b + \bd) - \sqrt{\kappa}\hat{\xi}_{1,2}. \label{Eq:EOMcav}
\end{gather}
\end{subequations}
Here, $\hat{\xi}_{1,2}$ represents standard Gaussian noise entering cavity $1,2$.
In this rotating frame, the dominant contributions of the mechanics on the cavity dynamics come from the low frequencies. 
Given the fast dynamics of the MR ($\Eb \gg \kappa, \tilde{g}$), one can solve the equation of motion of the MR adiabatically, i.e.~for $\dot{\b} = 0$. 
If we then substitute this solution into Eq.~\eqref{Eq:EOMcav} to eliminate $\b$ from the cavity equation of motion, 
one obtains an equation for the cavities that explicitly shows how the mechanical bath couples to the cavities. For $\gamma \ll \Eb$, the mechanical noise generates a contribution for $\dot{\hat{a}}_{1,2}$ of the form $\frac{\tilde{g}}{\Eb}\sqrt{\gamma}\hat{a}_{2,1}(\hat{\eta}_{\b} - \hat{\eta}_{\b}^\dag)$.
In a mean field approximation, one can then roughly approximate the rate $\Gamma_{i}$ at which the mechanics heats the cavity $i$, which reads:
\begin{equation}
	\Gamma_i \sim \gamma\frac{\tilde{g}^2}{\Eb^2}\bar{n}_j(2\nthb + 1) \approx \gamma\frac{g^2}{\Eb^2}\bar{n}_j(2\nthM + 1) e^{4r}. \label{Eq:Gamma}
\end{equation} 
Here $\bar{n}_j$ represents the mean number of photons in cavity $j = 2,1$ for $i = 1,2$. For a MR coupled to a thermal bath of temperature $\nthM$ and parametrically driven such that $e^{2r} \gg 1$, $\Gamma_i$ gives an estimate of the rate at which the cavity $i$, coupled to the MR via the single-photon coupling constant $g$, is heated. 

\subsection{Non-interacting mechanical Green's functions}

We calculate the non-interacting mechanical Green's function from the equations of motion \eqref{Eq:EOM}.
In terms of the $\chi[\omega]$ matrix, the retarded Green's functions are,
\begin{subequations}
\begin{gather}
	G^R_b[\omega] = \int_{-\infty}^{\infty}\mathrm{d}tG^R_b(t)e^{i\omega t} = -i\chi_{1,1}[\omega] = \frac{\cosh^2 r}{\omega - \Eb + i\gamma/2}
	- \frac{\sinh^2 r}{\omega + \Eb + i\gamma/2}, \\
	\tilde{G}^R_b[\omega]  = \int_{-\infty}^{\infty}\mathrm{d}t\tilde{G}^R_b(t)e^{i\omega t} = i\chi_{1,2}[\omega] = \frac{\sinh 2r}{2}\left(\frac{1}{\omega - \Eb + i\gamma/2} - \frac{1}{\omega + \Eb + i\gamma/2} \right).
\end{gather}
\end{subequations}
Without the parametric drive, the off-diagonal Green's function $\tilde{G}^R_b[\omega]$ vanishes. 
In the large $r$ limit ($e^{2r} \gg 1$) and for weak dissipation ($\gamma \ll \Eb$), $G^R_b[\omega] = \tilde{G}^R_b[\omega]$.

As we work in the interaction picture for the cavities, the shortest time scales relevant to the photons are $1/\kappa, 1/\tilde{g}$. Hence, the important contributions of the MR Green's functions to the effective two-photon interaction occur at low frequencies, i.e.~$\omega \ll \kappa, \tilde{g}$.
Consequently, to amplify the photon-photon interaction, one can decrease the parametric drive detuning $\Delta$ and increase its strength $\lambda$, which lowers the energy $\Eb$ and increases the exponential amplification $r$. However, to ensure that the interaction is sufficiently broadband, in other words local in time, one always needs to have $\Eb > \kappa, \tilde{g}$.

The non-interacting Keldysh Green's function of the MR are
\begin{subequations}
\begin{align}
	G^K_b[\omega] &= -i \gamma(2\nthM+1)(\vert\chi_{11}[\omega]\vert^2 + \vert\chi_{12}[\omega]\vert^2),\\
	\tilde{G}^K_b[\omega] &= -i \gamma(2\nthM+1)(\chi_{11}[\omega]\chi_{21}^*[\omega] + \chi_{12}[\omega]\chi_{22}^*[\omega]).
\end{align}
\end{subequations}
Their expressions are awkward, therefore we only focus on the limit $e^{2r} \gg 1$ and $\gamma \ll \Eb$. In these limits, one gets,
\begin{align}
	G^K_b[\omega] = \tilde{G}^K_b[\omega] = (1+2\nthM)\frac{e^{4r}}{8}\left( \frac{1}{\omega - \Eb + i\gamma/2} + \frac{1}{\omega + \Eb + i\gamma/2} - \frac{1}{\omega - \Eb - i\gamma/2} - \frac{1}{\omega + \Eb - i\gamma/2} \right).
\end{align}

As discussed above, The Keldysh Green functions capture the fact that the MR produces additional noises on the cavities. 
As in the case of the retarded Green functions, 
when $G^K_b[\omega]$ is peaked far from any relevant frequencies for the cavities dynamics, i.e.~$\Eb > \kappa, \tilde{g}$, the additional noise is off-resonant and scale as $\frac{\gamma}{\Eb}(1+2\nthM)e^{4r}$.

\section{Non-Gaussian cavity states}

\begin{figure}[t]
\centering
\includegraphics[width=0.66\textwidth]{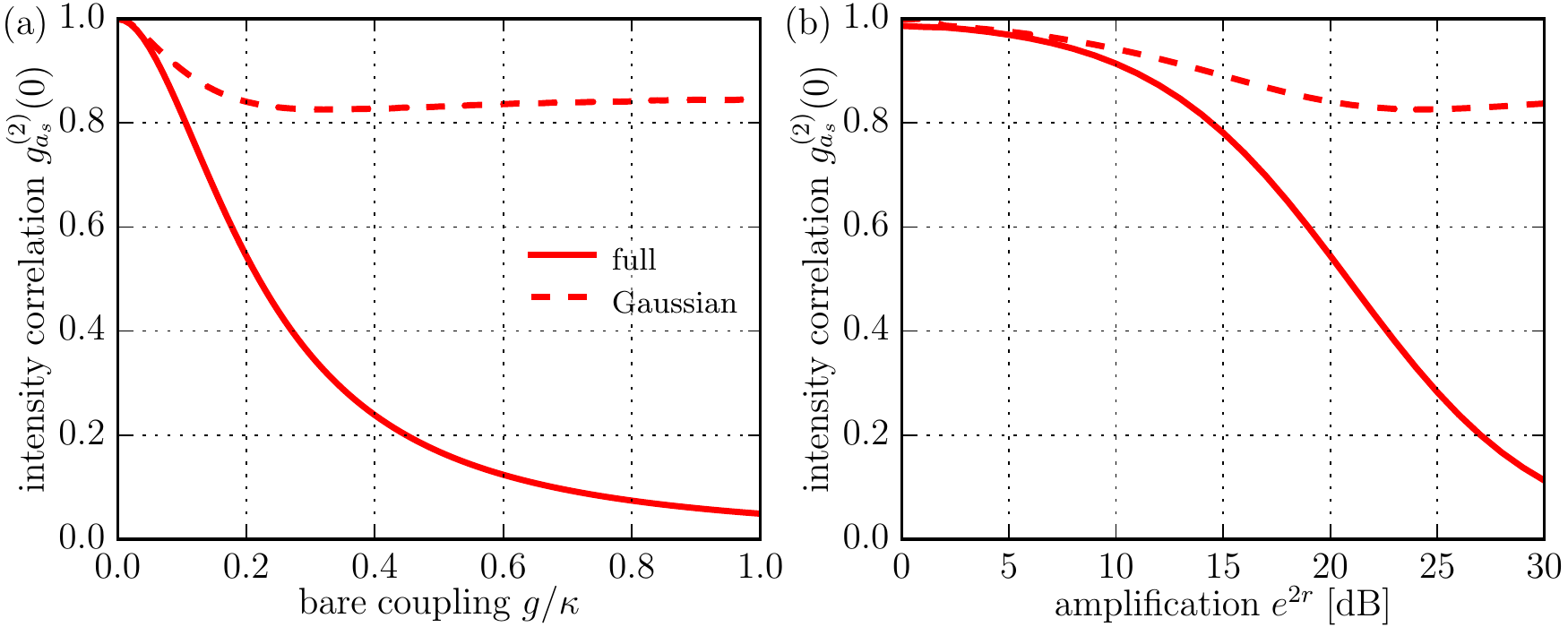}
\caption{Comparison between the $g_{a_s}^{(2)}(0)$ correlation function for Hamiltonian $\hat{H}_\mathrm{SRP}$ with the parameters of Fig.~3 of the main text (full lines) and the intensity correlation supposing Gaussian optical states (dashed lines).
The Gaussian result only involves the expectation values $\langle \hat{a}_s \rangle$, $\langle \hat{a}_s^\dag\hat{a}_s \rangle$ and $\langle \hat{a}_s^2 \rangle$.
The clear different indicates that the enhanced photon-photon interaction induces non-Gaussian photonic states.
This is reminiscent of true photon antibunching due to the nonlinear interaction.}
\label{FigGauss}
\end{figure}

The non-Gaussian character of the cavity state is addressed in Fig.~\ref{FigGauss}, where the $g_{a_s}^{(2)}(0)$ correlation functions plotted in Fig.~3 of the main text are compared to the result that would be obtained if the states were Gaussian. 
In the case of purely Gaussian states, it is possible to calculate $g_{a_s}^{(2)}(0)$ solely in terms of the covariance matrix of the cavity state, i.e.~by using Wick's theorem~\cite{Lemonde_PRA_2014}.
The results for the Gaussian state presented in Fig.~\ref{FigGauss} thus consist of calculating the covariance matrix of the optical state and extrapolating the corresponding $g_{a_s}^{(2)}(0)$ via a Wick expansion of the term $\langle \hat{a}_s^{\dag 2}\hat{a}_s^{2} \rangle$. 
For increasing $g$, the result for the Gaussian state deviates from the full solution, thus explicitly demonstrating that the photons are in a non-Gaussian state, and that the $g_{a_s}^{(2)}(0)$ suppression is due to true photon blockade. As expected, as the nonlinearity vanishes $g \rightarrow 0$, the optical state becomes purely Gaussian.

\section{Emission of cavity states to input/output waveguide}
\label{SecWigner}

\begin{figure}[b]
\centering
\includegraphics[width=0.66\textwidth]{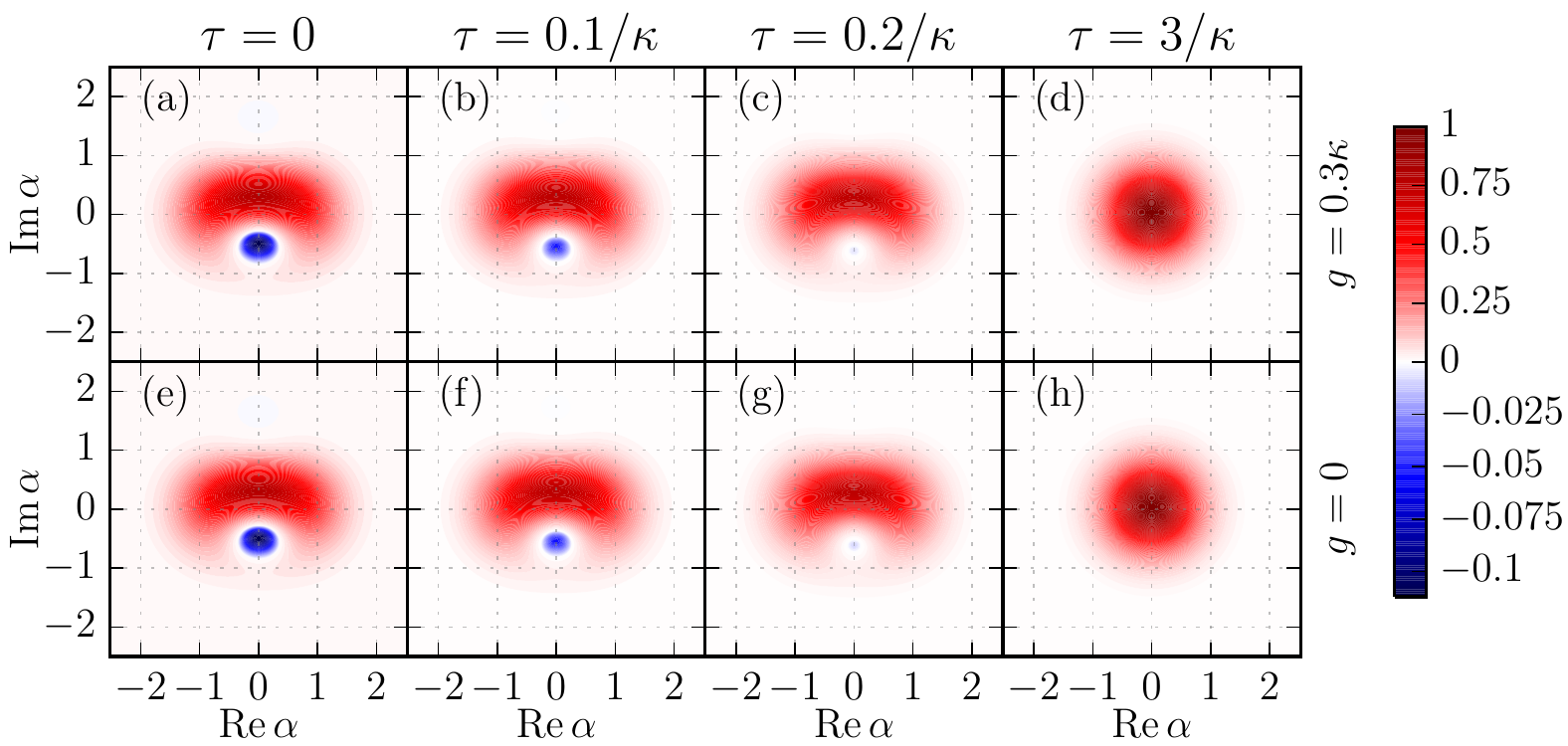}
\caption{Decay dynamics of a quantum state after the parametric drive is switched off with the TD scheme.
The initial state corresponds to the Wigner function plotted in Fig.~4~(c) of the main text [panels (a) and (e) here].
The decay of this initial state in cavity~1 in presence of the residual optomechanical coupling $g=0.3\kappa$ (c-d) is compared to the non-interacting cavity $g=0$ (f-h) at different times.
The two dynamics are identical, ensuring that the quantum state is perfectly transferred to propagating photons.
}
\label{Figdecay}
\end{figure}

The protocol to prepare negative optical Wigner functions explained in the main text shows the intracavity dynamics.
As discussed, by turning off the parametric drive rapidly using a transitionless driving (TD) scheme, the optomechanical interaction is effectively turned off, and the intracavity state is converted into a propagating state.
We show here that once the parametric drive is turned off, the remaining weak optomechanical coupling $g$ plays no role in the dynamics (both because it is not enhanced by the coupling, and because it is now non-resonant).
The Hamiltonian is then given by Eq.~(1) of the main text with $\lambda=0$,  
\begin{equation}
\hat{H} = \Delta \hat{b}^\dag\hat{b} + g[ \ao\atd\hat{b}+ \aod\at\hat{b}^\dag].
\label{Hlambda0}
\end{equation}
The nonlinear interaction is weak ($g$ instead of $\tilde{g}$) and strongly off resonant ($\Delta$ instead of $E_\beta\ll\Delta$).
Moreover, in the absence of excitation for both the MR and cavity 2, the interaction is suppressed,
thereby preventing the residual nonlinear interaction from affecting the dynamics of the cavity-1 state.
Finally, the non-rotating wave term that is neglected in Eq.~\eqref{Hlambda0} (involving $\atd\ao\hat{b}^\dag$) remains negligible as it is highly non-resonant [see text after Eq.~\eqref{HRWA2}].

The decay dynamics is shown in Fig.~\ref{Figdecay} for $g=0.3\kappa$ and compared to the decay of the non-interacting cavity (i.e.~for $g=0$).
The Wigner functions are identical and the fidelity between the two evolutions stays at 1.
This agreement ensures that the initial quantum state is perfectly transferred to the propagating photons of the outgoing field and can be sent to remote quantum systems.

\section{Parametrically driving the cavity}

\begin{figure}[t]
  \centering
  \includegraphics[width=0.5\textwidth]{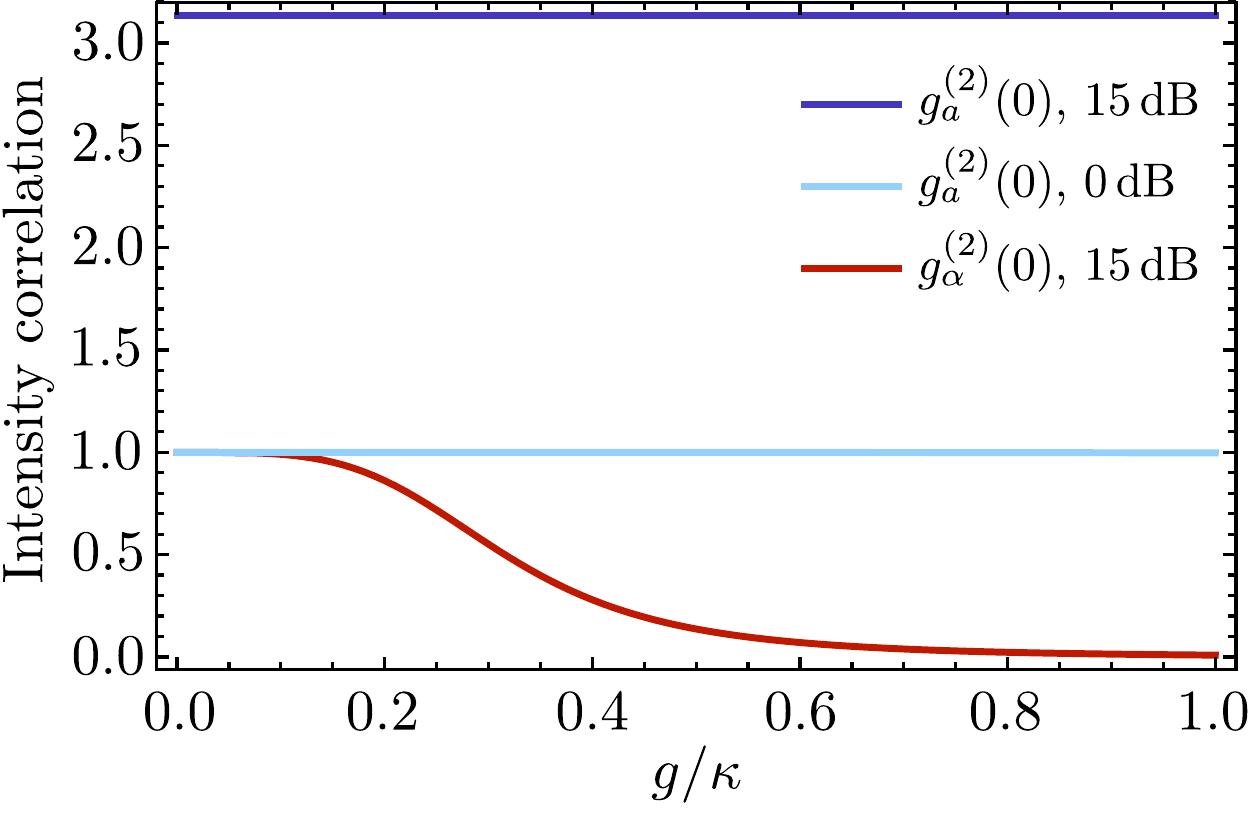}
  \caption{Intensity correlation functions of the parametrically driven cavity mode as a function of the OM coupling $g$. We compare the $g^{(2)}(0)$ function of the $\alpha$ mode (red line) to the real photon $\hat{a}$ (dark blue) for an amplification of $15\,\mathrm{dB}$. We see that while $g^{(2)}(0)$ goes below $1$ for the $\alpha$ mode, there is no suppression for the real photons. More precisely, the parametric drive increases the intensity correlation of the real photons, as can be seen by comparing to $g^{(2)}_a(0)$ without parametric drive (light blue line). The parameters used are $\omega_M = 50\kappa$, $\omega_{d} = -g^2\cosh^2(2r_c)/\omega_M$, $\epsilon = 0.001\kappa$, $\gamma = 10^{-4}\kappa$ and zero temperature dissipative baths for the MR and the $\alpha$ mode.
} 
 \label{Fig:g2Nori}
\end{figure}

In our work, we consider how parametric mechanical driving enhances the interaction in an OM system.  The dual situation was studied recently by Xin-You L\"{u} \textit{et al.} in Ref.~\onlinecite{Liu_Nori_PRL_2015}, where the cavity in an OM system is parametrically driven. 
Because the OM interaction [$\propto\hat{a}^\dag\hat{a}(\hat{b}^\dag+\hat{b})$, see Eq.~\eqref{Hsqueezedcavity}] is fundamentally asymmetric between photons and phonons, the photons, unlike the phonons, do not mediate any effective interaction.
Parametrically driving the cavity thus gives rise to a different physics and, in particular, does not lead to the enhancement of the nonlinear interaction at the single-photon level. 
We explain this point in more details here, and show explicitly that the approach of Ref.~\onlinecite{Liu_Nori_PRL_2015} does not result in a photonic intensity correlation function satisfying $g^{(2)}(0) < 1$.

For simplicity, we consider a single cavity mode coupled to the MR as it is studied in Ref.~\onlinecite{Liu_Nori_PRL_2015}. 
The corresponding coherent dynamics is governed by the following Hamiltonian,
\begin{align}
	\hat{H} = \Delta \hat{a}^\dag\hat{a} - \tfrac{1}{2}\lambda( \hat{a}\hat{a}+ \hat{a}^\dag\hat{a}^\dag) + \omega_M \hat{b}^\dag\hat{b} + g\hat{a}^\dag\hat{a}( \hat{b} + \hat{b}^\dag).
\label{Hsqueezedcavity}
\end{align}
Here $\Delta = \omega_\cav - \omega_p$ with $\omega_\cav$, $2\omega_p$ being respectively the cavity and the parametric drive frequencies and $\lambda$ is the parametric drive strength. 
We treat the squeezed photons as we treated the parametrically driven phonons in the text.
We diagonalize the quadratic part of $\hat{H}$ by applying the Bogoliubov transformation $\hat{a} = \cosh r\hat{\alpha} + \sinh r\hat{\alpha}^\dag$ with 
$\tanh 2r = \lambda/\Delta$. 
In this squeezed basis, the energy of the $\alpha$-mode is $E_\alpha = \Delta/\cosh(2r_\cav)$ and the Hamiltonian reads,
\begin{equation}
	\hat{H} = E_\alpha \hat{\alpha}^\dag\hat{\alpha} + \omega_M \hat{b}^\dag\hat{b} +  g\cosh (2r_\cav) \hat{\alpha}^\dag\hat{\alpha}(\hat{b}^\dag + \hat{b}) + \tfrac{1}{2}g\sinh (2r_\cav)( \hat{\alpha}\hat{\alpha} + \hat{\alpha}^\dag\hat{\alpha}^\dag)(\hat{b}^\dag + \hat{b}).
\label{Eq:HSqCav}
\end{equation}

As in Ref.~\onlinecite{Liu_Nori_PRL_2015}, we focus on the limit $E_\alpha \gg \omega_M, g\sinh(2r_\cav)$ and neglect the off-resonant nonlinear interaction [last term of Eq.~\eqref{Eq:HSqCav}].
After eliminating the mechanical degree of freedom with a polaron transformation $\hat{U} = \exp [g\cosh (2r_\cav)(\hat{b}^\dag - \hat{b})\hat{\alpha}^\dag\hat{\alpha}/\omega_M]$, the OM interaction generates a Kerr nonlinearity for the $\alpha$ mode of the form 
\begin{equation}
	\hat{H}_\mathrm{pol} = \hat{U}\hat{H}\hat{U}^\dag = E_\alpha \hat{\alpha}^\dag\hat{\alpha} + \omega_M \hat{b}^\dag\hat{b} -  K_\cav (\hat{\alpha}^\dag\hat{\alpha})^2, \label{Eq:Hpolalpha}
\end{equation}
with $K_\cav=g^2\cosh^2 (2r_\cav)/\omega_M$.
It thus follows that photonic parametric driving leads to an extremely nonlinear energy spectrum (similar to our approach).  However, the eigenstates corresponding to the nonlinearity in Eq.~\eqref{Eq:Hpolalpha} are \textit{not few photon states}; they are $\alpha$-mode Fock states, and correspond to squeezed photonic Fock state.  They thus necessarily involve extremely large photon numbers when $r_\cav \gg 1$.  Thus, to make use of the nonlinearity in Eq.~\eqref{Eq:Hpolalpha} one must necessarily work with states with large photon number.  This is in stark contrast to our scheme, where the enhanced nonlinear spectrum corresponds to states having only a few photons.

This difference in the eigenstates associated with the enhanced nonlinearity is not just a question of semantics: it leads to crucial observable differences.  As an example, we consider again the $g^{(2)}(0)$ intensity correlation function of the cavity photons. 
As in Ref.~\onlinecite{Liu_Nori_PRL_2015}, we consider a weak drive of the form $\hat{H}_\mathrm{drive} = \epsilon (\hat{a} e^{i\omega_d t} + \mathrm{H.c.})$, with frequency $\omega_d$ in the frame rotating at $\omega_p$, to probe the intensity fluctuations of the cavity. 
For a very weak drive ($\epsilon \ll g$) that is near resonance with the Bogoliubov mode (small detuning $\delta_\alpha = E_\alpha - \omega_d$), the drive term reduces to $\hat{H}_\mathrm{drive} \approx \epsilon \cosh r(\hat{\alpha} e^{i\omega_d t} + \mathrm{H.c.})$. 
In the frame rotating at $\omega_d + \omega_p$, one gets the following final Hamiltonian,
\begin{equation}
	\hat{H} =  \delta_\alpha \hat{\alpha}^\dag\hat{\alpha} + \omega_M \hat{b}^\dag\hat{b} +  g\cosh (2r_\cav) \hat{\alpha}^\dag\hat{\alpha}(\hat{b}^\dag + \hat{b}) + \epsilon \cosh r_\cav(\hat{\alpha} + \hat{\alpha}^\dag). \label{Eq:HSqCav2}
\end{equation}

We use a standard Lindblad master equation to calculate the equal-time intensity correlation function of the cavity under the dynamics of the Hamiltonian Eq.~\eqref{Eq:HSqCav2} and we consider that both the mechanical mode and the $\alpha$ mode are coupled to zero temperature baths. Zero temperature dissipation for the $\alpha$ mode is possible if the cavity is also driven by squeezed light with a properly tuned phase~\cite{Liu_Nori_PRL_2015}. 

In Fig.~\ref{Fig:g2Nori}, we compare the real photon intensity correlation $g^{(2)}_{a}(0)$ to the $\alpha$ mode intensity correlation $g^{(2)}_{\alpha}(0)$ as a function of $g$ for an amplification of $15\,\mathrm{dB}$ [i.e.~$e^{2r} \simeq 32$]. As expected from the Kerr interaction in Eq.~\eqref{Eq:Hpolalpha}, $g^{(2)}_{\alpha}(0)$ drops rapidly below $1$ as $g$ is increased. 
However, it is not the case for the real photons ($\hat{a}$):
 even with no additional coherent drive, the large photon population in the cavity induced by the parametric drive would yield $g^{(2)}_{a}(0) \sim 3$.
Physically speaking, the effects of the enhanced nonlinear interaction on the cavity are sitting on top of a large photon-number Gaussian state, making them both hard to detect and exploit. This is again in stark contrast to our scheme, where there is no large background number of photons obscuring the interesting physics.
Furthermore, the relevant observable directly measured by a photodetector is $g_a^{2}(0)$, not the $g_\alpha^{2}(0)$ intensity correlation of the $\alpha$ mode.

Finally, it is worth noting that the scheme of Ref.~\onlinecite{Liu_Nori_PRL_2015} would apply to any system having a weak Kerr interaction, and does not make use of any special aspect of an optomechanical system; parametrically driven Kerr cavities have been studied far earlier in the literature, see e.g. Ref.~\onlinecite{Wielinga_1994}.  On a heuristic level, parametrically driving the cavity puts it in a state with a large number of photons in it; given this large population, it is not surprising that any intrinsic weak nonlinearity will play a larger role.

\bibliographystyle{apsrev}
\bibliography{MALPapersRefs}